\documentstyle[aps,epsf,twocolumn,prl]{revtex}

\newcommand{\beq}{\begin{equation}} \newcommand{\eeq}{\end{equation}}

\author{Christophe Josserand \thanks{present address: Laboratoire de
Mod\'elisation en M\'ecanique, CNRS UMR 7607,  
8 Rue du Capitaine Scott, 75015 Paris}, Alexei Tkachenko\thanks{ 
present address: Bell Labs, 600 Mountain Ave., Rm 1D329, Murray Hill, 
NJ 07974}, Daniel M. Mueth, and  Heinrich M. Jaeger\\ 
\it The James Franck Institute, \\ 
\it The University of Chicago, Chicago, Illinois 60637 } 
\title{\bf Memory Effects in Granular Material }  
 
\begin{document}   
\wideabs{
\maketitle   
\begin{abstract}   
We present a combined experimental and theoretical study of memory   
 effects in  vibration-induced compaction of granular materials.  In   
 particular, the response of the system to an abrupt change in shaking   
 intensity is measured. At short times after the perturbation a   
 granular analog of aging in glasses is observed. Using a simple   
 two-state model, we are able to explain this short-time response.   
 We also discuss the possibility for  the system to obey  an   
 approximate pseudo-fluctuation-dissipation theorem relationship and   
 relate our work to earlier experimental and theoretical  studies of   
 the problem.   
   
   
\end{abstract}   
}
   
Granular materials comprise an  important class  of complex systems whose 
simple fundamental mechanics gives rise to rich   macroscopic 
phenomenology~\cite{JNB}. Recent experiments on granular 
compaction~\cite{comp1,comp2} suggest they are an ideal   system for 
studying jamming, a phenomenon lying outside the   domain  of 
conventional statistical physics, yet highly reminiscent   of 
glassiness.  These studies showed that a loose packing of glass   beads 
subjected to vertical ``tapping'' slowly compacts, asymptoting to a 
higher steady state packing fraction.  This ``equilibrium'' packing 
fraction  is somewhat lower than the random close packing limit, 
$\rho_{\rm rcp} \approx 0.64$, and  is a decreasing function of the 
vibration intensity, typically parameterized by $\Gamma$, the peak 
applied   acceleration normalized by gravity, $\rm{g}$. 
The relaxation dynamics are extremely slow, taking  many thousands of 
taps for the packing fraction, $\rho$, to approach its steady state 
value. During this evolution, $\rho$ increases logarithmically with 
the number of taps, $t$, which is typical for self--inhibiting 
processes~\cite{boutreux}.  The average time scale $\tau$ of 
the relaxation decreases with $\Gamma$, and in this sense the shaking 
intensity plays, at least qualitatively, the role of temperature. For 
small $\Gamma$, the relaxation rate becomes so  slow that the system 
cannot   reach the steady state density within  the experimental time 
scale. It was also found that compaction can be maximized through an 
annealing procedure.  This process  involves a slow ``cooling'' of the 
system starting from a high shaking intensity $\Gamma$.  These slow 
relaxation and annealing properties of this system  are reminiscent of 
conventional glasses.  Another qualitative similarity to glasses is 
observable in  the density fluctuation spectrum of the granular system 
near equilibrium. The spectrum  was found to be strongly 
non-Lorentzian~\cite{comp2}, revealing the existence of multiple time 
scales in the system. The shortest and the longest relaxation 
timescales differ by as much as  three order of magnitude, and the 
behavior of the spectrum for the intermediate frequencies is highly 
non-trivial; in certain regimes it can be fitted with a  power 
law.  
 
These previous experimental observations are suggestive of glassy
behavior and this connection has been explored in recent models of
compaction using ideas from magnetic systems~\cite{nicodemi}. However,
a  more direct test of the glassy nature of granular compaction comes
from measurements of the response of the system to sudden
perturbations in the effective temperature, given by $\Gamma$. This
idea originates from classical  experiments for the study of aging in
glasses~\cite{glass} and has recently been explored using computer
simulations~\cite{FDT1}.   In this letter we present direct
experimental observations  of memory effects in a vibrated granular
system obtained by measuring  the short-time response to an
instantaneous change in tapping  acceleration $\Gamma$ and propose a
simple theoretical framework.
   
We used  the experimental set-up described in 
refs.~\cite{comp1,comp2}:~1~mm--diameter  glass beads were vertically 
shaken in a tall, evacuated, 19~mm-diameter glass tube, and the 
packing density of the beads was measured using capacitors mounted at 
four heights along the  column. 
  
\begin{figure}[h]
\centerline{ \epsfxsize 8truecm \epsfbox{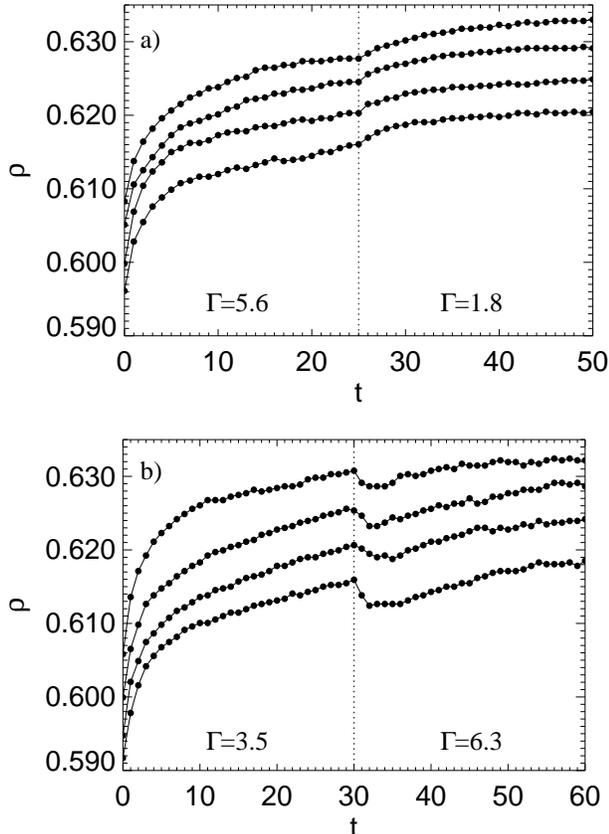} } 
\caption{\protect\small Evolution of the packing fraction, $\rho$, at 
four   heights in the column, as a function of tap number, $t$. Two 
different single-switch experiments:   (a) $\Gamma$  was lowered from 
$5.6$ to $1.8$ at $t_0= 25$ ; and (b)   $\Gamma$  was increased from 
$3.5$ to $6.3$ at $t_0=30$. Curves are shifted vertically for 
clarity. Each   curve is an average over $4$ runs, and   the 
measurement uncertainty in $\rho$ is $4\times   10^{-4}$. 
\label{evolu}}   
\end{figure}

The simplest form of this experiment consists of a single
instantaneous change of vibration intensity from $\Gamma_1$ to
$\Gamma_2$ after $t_0$ taps.  For $\Gamma_2<\Gamma_1$
(Fig.~\ref{evolu}a) we found that on short time scales the
compaction rate increases.  This is in sharp contrast to what one may
expect from the long-time behavior found in previous experiments where
the relaxation is slower for smaller vibration accelerations.  For
$\Gamma_2>\Gamma_1$ (Fig.~\ref{evolu}b) we found that the system
dilates immediately following $t_0$. 

 These results too, are opposite
from the long-time behavior seen in  previous experiments where the
compaction rate increased: Not only does the compaction rate decrease,
it becomes negative (i.e. the system dilates). Note that after
several taps the ``anomalous'' dilation ceases and there is a
crossover to the ``normal'' behavior, with the relaxation  rate
becoming the same as in constant--$\Gamma$ mode. Thus, most
of the shaking history is forgotten after a short time.

\begin{figure}[h]   
\centerline{ \epsfxsize=8truecm \epsfbox{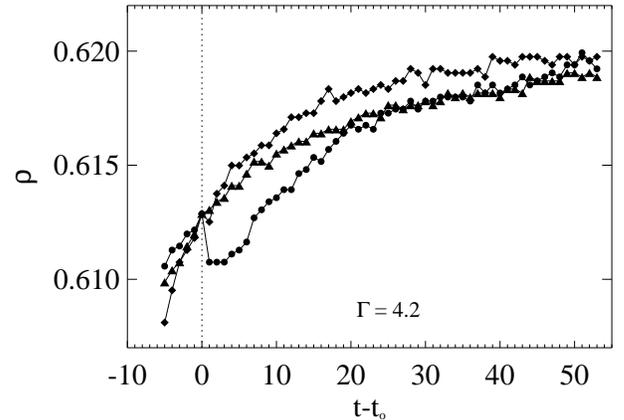}  }   
\caption{\protect\small The time evolution of packing fraction $\rho$ 
for a system which was compacted to $\rho_0=0.613$ at time $t_0$ 
using three different accelerations: $\Gamma_1=1.8$ (circles), 
$\Gamma_0=4.2$ (triangles), and   $\Gamma_2=6.3$ (diamonds).  After 
the density $\rho_0$ was   achieved, the system was vibrated at 
acceleration $\Gamma_0$. The   evolution for $t>t_0$ depended strongly 
on the pre-history. Each   curve is an average over four experimental 
runs.  }\label{fancy} 
\end{figure}

These data constitute a short-term memory effect: the future evolution
of $\rho$ after time $t_0$ depends not only on $\rho(t_0)$, but also
information about the previous tapping history, contained in other
``hidden'' variables.  In order to demonstrate this in a more explicit manner, 
we modified the above experiment.  In this second set of three 
experiments the systems was driven to {\em the same density} $\rho_0$ 
with three different accelerations $\Gamma_0$, $\Gamma_1$, and 
$\Gamma_2$. After $\rho_0$ was achieved at time $t_0$, the system 
was tapped with {\em the same intensity} $\Gamma_0$ for all three 
experiments. As seen in Figure~\ref{fancy}, the 
evolution for $t>t_0$ strongly depends on the pre-history. The need 
for extra state variables in the problem is consistent with strongly 
non-Lorentzian behavior of the fluctuation spectrum, observed in 
earlier experiments\cite{comp2}. Indeed, if the evolution could be 
prescribed by a single master equation for local density, it would 
result in a single-relaxation-time exponential decay of the density 
fluctuations near equilibrium. Instead, a wide distribution of 
characteristic times is suggested by the spectrum. 
   
To give a theoretical interpretation of the above results, we
view the problem as an evolution in the space of discrete
``microscopic'' states corresponding to different realizations of the
packing topology (i.e. of the contact network).  For each tap there is a
possibility for a transition from one microscopic state to
another. Since the dynamics is dissipative and the system is under
external gravity, a transition to a denser configuration is typically
more probable that the reverse one. We now assume that the short--term
dynamics  of the system are dominated by a number of local flip--flop
modes with relatively high transition rates in both directions.

\begin{figure}[h]   
\centerline{ \epsfxsize=6truecm \epsfbox{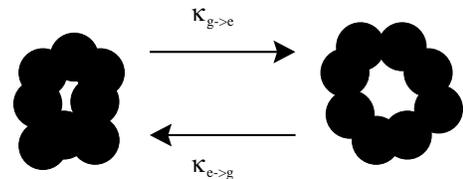}  } 
\caption{\protect\small Schematic description of the two-state model: 
at every tap a transition can occur from the ground to the excited 
state or vice versa. 
\label{schema}}   
\end{figure}

This model replaces the complicated configuration space with  a
set of independent two-state systems, each of which is characterized
by two transition rates, $\kappa_{e\rightarrow g} > \kappa_{g
\rightarrow e}$. $\kappa_{e\rightarrow g}/ \kappa_{g \rightarrow e}$
gives  the ratio of the equilibrium  probabilities of populating each
state:    ``ground'' and ``excited''.  As we have argued, the higher
probability  ground state is typically the one with higher density,
i.e. the volume change  $ v$ between the ground  and the excited
states is normally positive (see Fig.~\ref{schema} for a schematic
description of the model). Our  two-state approximation is close in
its spirit  to  recent  Grinev--Edwards and  de Gennes
models~\cite{GrEd,Deg}.

We now introduce the concept of a base-line density, $\rho_b$, which
corresponds to  all the elementary modes at their ground states.
Obviously, the experimentally--observed density is different from
$\rho_b$ due to a non-zero fraction of excited states: \beq
\rho=\rho_b(t)\left(1-\frac{1}{V}\sum_n
v^{(n)}\left(1+\frac{\kappa^{(n)}_{g \rightarrow e}}
{\kappa^{(n)}_{e\rightarrow g}}\right)^{-1}\right).  
\eeq 

The summation here  is performed over all  the dominant two-state modes,
$V$ is the total volume and $v^{(n)}$ is the volume difference between
the excited and the ground  $n$-th state.  Assuming that the vibration
intensity $\Gamma$ is a qualitative analog of temperature, we expect
the population of the excited states, $P(\Gamma)=(1+{\kappa^{(n)}_{g
\rightarrow e}}/{\kappa^{(n)}_{e\rightarrow g}})^{-1}$, to grow with
$\Gamma$, starting  from  zero at $\Gamma=0$. Hence, for a given
$\rho_b$, the total  density $\rho$ will be lower at higher
acceleration.  This explains the observed effect of an abrupt change
of  $\Gamma$. After a switch from  $\Gamma_1$ to $\Gamma_2$ at time
$t_0=0$, the flip-flop mode contribution to the total density would
relax to its new equilibrium value in the following way:
   
\beq
\label{Delta}   
\Delta_{\Gamma_1\Gamma_2}(t)=\rho_b \int { v F_{\Gamma_1,\Gamma_2}(v,
\kappa)\left(1-exp   (-\kappa t)\right)d v d \kappa } \eeq Here
$\kappa$ is the   relaxation rate of an individual mode, and the
distribution function   $F_{\Gamma_1,\Gamma_2}( v, \kappa)$ is
introduced as follows:
   
\begin{eqnarray}   
F_{\Gamma_1,\Gamma_2}( v, \kappa)\equiv \frac{1}{V} \sum_n
\left(P^{(n)}(\Gamma_2)-P^{(n)}(\Gamma_1)\right)\delta(v-v^{(n)})
\nonumber \\   \delta\left(\kappa-\kappa^{(n)}_{g \rightarrow
e}(\Gamma_2)-\kappa^{(n)}_{e\rightarrow g}(\Gamma_2)\right).
\label{F}   
\end{eqnarray}   
   
Since the observed density changes in compaction experiment are normally less 
than $1\%$ of the total density\cite{comp1,comp2}, one can estimate the 
typical separation between
neighboring  flip-flop systems as 5 particle sizes, which yields a
good   justification for  our no-coupling approximation. According to
Eq.~(\ref{Delta}), if $F_{\Gamma_1,\Gamma_2}$  does not vanish in the
limit   $\kappa\rightarrow 0$, the late stage of the relaxation of
$\Delta_{\Gamma_1\Gamma_2}(t)$ is   given  by the power law:

\begin{eqnarray}   
\label{miracle}   
\delta_{\Gamma_1,\Gamma_2}(t)\equiv\Delta_{\Gamma_1\Gamma_2}(t)-\Delta_{\Gamma_1\Gamma_2}(\infty)=
\\ \rho_b  \int {v F_{\Gamma_1,\Gamma_2}(v, \kappa)e^{-\kappa t} d  v
d \kappa }   \sim\frac{1}{t}.  \nonumber
\end{eqnarray}   
   
Note that $\rho_b$ is also dependent on time: although this cannot be
described  within  our  two-state approximation, the collection of
elementary modes  slowly  evolves. Thus,  one can observe two
different processes: on short time scales, a fast relaxation due to
the flip--flop modes is dominant, while over the long times, the dynamics
are determined by the logarithmically slow evolution  of the baseline
density $\rho_b(t)$. The  crossover between the two regimes is
particularly obvious in  Fig.~\ref{evolu}b, where it results in a
non-monotonic   evolution. For experiments performed at
sufficiently   late stages of the  density  relaxation, the
dynamics of the   baseline density  could be neglected  compared to
the contribution of   the  flip-flop  modes (note that what we call
a late-stage   relaxation corresponds in fact
to  mesoscopic   time scales which  are always shorter than the
relaxation time  for $\rho_b$).  It has  to be emphasized that the
described experiments   provide us with  a  tool  for study of the
response of the system,   which is {\em not  limited} to the
nearly--equilibrium regime.
   
One can use our simple model to predict the response of the system  to
a more  complicated pattern of changes of $\Gamma$. First, we reach, 
using annealing dynamics, a ``quasi-steady'' state at amplitude 
$\Gamma_0$, so that one can consider $\rho_b$ constant later on.
 Let us switch  the
shaking  acceleration  from $\Gamma_0$ to $\Gamma_1$ for a  finite
number of taps  $\delta t$, and then switch it  back to  $\Gamma_0$.
During the intermediate $\Gamma_1$--stage, the system  does not have
enough time  to completely relax to its new  equilibrium. In our
two-state model, the modes whose  relaxation rate (at $\Gamma_1$) is
below $\delta t^{-1}$ remain  unrelaxed. Assuming that  the slow
modes at $\Gamma_1$ are  mostly the same as at $\Gamma_0$, we can
calculate  the  backward  density relaxation similarly to
Eq.~(\ref{miracle}), with  $F(v, \kappa)$  effectively depleted below
a minimal rate,  $\kappa_0$. This  cut-off  frequency, $\kappa_0$, is
expected to  decrease monotonically  with increasing  perturbation
duration  $\delta  t$. The resulting density relaxation after
returning to $\Gamma_0$ is given by:
    
\beq \delta_{\Gamma_1,\Gamma_0}(t) \sim\frac{exp(-\kappa_0 t)}{t}.
\label{miracle1}   
\eeq

\begin{figure}[t]
\centerline{ \epsfxsize=8truecm \epsfbox{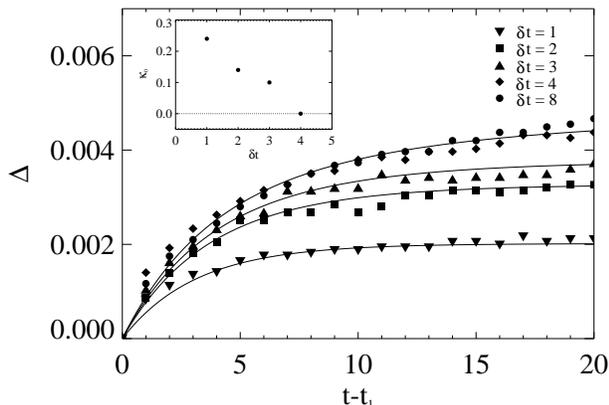}  } 
\caption{\protect\small The density relaxation
($\Delta=\rho(t)-\rho(t_1)$) of the system, prepared by tapping for a
long time  at $\Gamma_0=1.8$ and then tapping for a
variable number, $\delta t$, of taps at a ``hotter'' intensity
$\Gamma_1=4.2$ before being returned to $\Gamma_0$ at time $t_1$.
Note that $\rho(t_1)$ is a function of $\delta t$.  Thus, while the
asymptotic value of $\Delta$ depends on $\delta t$, all curves asymptote
to the same density $\rho(\infty)$, which depends only on $\Gamma_0$
and not on $\delta t$. The
solid lines represent the theoretical curves, with  appropriate
values of the parameter $\kappa_0$. The dependence of the  cut-off
rate $\kappa_0$ on the waiting time $\delta t$ is shown at  the insert
for $\delta t \le 4$ taps. We do not show the  value for $\delta t =8$
since we found it null within the error bar,  as for   $\delta t
=4$. Each  experimental graph is an average of $12$ runs.
\label{ageing}}   
\end{figure}

We tested these predictions by performing this three stage experiment,
varying the duration, $\delta t$, of the perturbation($\Gamma_1$)
stage~(Fig.~\ref{ageing}). As predicted, the  time needed to  recover 
the steady-state density increases with the number of taps $\delta t$
spent in the ``hot'' regime $\Gamma_1 >\Gamma_0$. In the coordinates
chosen, the relaxation curves should follow  the $\delta t=\infty$
dynamics until the saturation at the  cut-off time,
$\kappa_0^{-1}(\delta t)$.
We approximate the distribution  function $F$ by a constant above this
low frequency cut-off at $\kappa_0^{-1}(\delta t)$, up to a
high-frequency cut-off, $\kappa_h  \simeq 1 \rm{tap}^{-1}$. This
eliminates the unphysical low-$t$  divergence in
Eq.~(\ref{miracle1}). Figure~\ref{ageing} shows fits of the data to
Eq.~(\ref{Delta}), where
$\kappa_0(\delta t)$ is determined from the fit. The best-fit is achieved at
$\kappa_h=0.4$, and the  variation of this parameter would result in a
simple rescaling of the time  axis.

Figure~{\ref{ageing}}
demonstrates good agreement between model and experiment, with some
systematic error at the earliest relaxation stage  (which is an
expected result of our oversimplified  description of the short time
dynamics). For the late stage relaxation, we conclude that
(i) within our experimental precision, the $\delta
t=\infty$  relaxation is consistent with  the predicted $1/t$ law;
(ii) finite--$\delta t$ relaxation curves can be parameterized by a
low frequency cut-off, $\kappa_0$;  and (iii) $\kappa_0$ is a decreasing
function of the waiting time $\delta t$, shown in the insert of the
Figure~{\ref{ageing}}.
We now relate  our picture to previous experimental and theoretical 
results.  As discussed earlier, the wide range of 
relaxation times  reveals itself both in our response measurements and 
in the the fluctuation  spectra of the density. It is tempting 
to relate these two kinds of data  through an analog of a 
Fluctuation-Dissipation Theorem (FDT). Of course, there  is no 
fundamental reason for FDT to be applicable to the granular 
system~\cite{glassfootnote}. Even  though the above two-state model 
could be mapped onto a   thermal system (in  which FDT is expected to 
work), the thermodynamic   variable conjugate to  density in the 
context of such a mapping  has   no clear physical 
meaning. Nevertheless, below we outline the   pseudo-FDT relationship 
expected for the  granular system under rather   natural 
approximation. Namely, we neglect the  correlation between the 
volume change $v$ and the life time $\kappa^{-1}$ of  an individual 
mode, i.e. assume $F_{0,\Gamma}(v, \kappa)= f(\kappa)g(v)$. Then the 
density   autocorrelation function can be written as follows: 
   
\begin{eqnarray}   
\langle\delta\rho(0)\delta\rho(t)\rangle_\Gamma=\frac{\rho^2}{2V}\int{\left(\langle   
v^2\rangle-\langle v\rangle^2\right)exp(-\kappa t)f(\kappa)d\kappa}=   
\nonumber \\ \rho \frac {\langle v^2\rangle-\langle   
v\rangle^2}{2\langle v \rangle V}\delta_{0,\Gamma}(t).   
\label{FDT}   
\end{eqnarray}   
   
Thus, the density correlator is simply proportional to the response   
function  corresponding to the switch between a very low acceleration   
(at which virtually  all the modes are in their ground states) and the   
given one, $\Gamma$. An  experimental check of this relationship   
requires further high-precision  studies of both the relaxation   
dynamics and the fluctuation spectrum.   

Our model also  gives a simple interpretation to the decreasing   
dependence of  the steady-state density on $\Gamma$: it can be   
attributed to the growth of the  population of the excited states,   
$P(\Gamma)$. Indeed, the corresponding  correction to the total   
density is about $1\%$, i.e. of the same order as the  variation of   
the equilibrium packing fraction with $\Gamma$~\cite{comp2}.   
     
The slow dynamics associated with the evolution of the base-line   
density can  also be addressed within our approach. For doing so we   
need to account for the  coupling of individual modes. Namely, it   
would be a reasonable assumption that  a relaxation of one mode to its   
ground state may frustrate such a transition  for some of its   
neighbors (e.g. in 3D the most compact local cluster can be  created   
only at the expense of less dense neighboring regions). Thus, we   
arrive  at an effective anti-ferromagnetic (AF) coupling (of an   
infinite strength)  between the two-state modes. This extension of our   
model  makes it remarkably  similar to the so-called reversible   
Parking Lot Model (PLM)~\cite{eli,alex},  which has been
 successful in  describing many aspects of granular   
compaction experiments~\cite{comp1,comp2}.   Recent simulations
based on the ``tetris model''~\cite{glass} for compaction also find slow glassy
responses to changes in $\Gamma$, but do not capture the short-term
memory effect described here.~\cite{NICO}.
  
In the PLM, D-dimensional space (the parking lot) gets packed with
finite-size  objects (cars), which may arrive and depart with fixed
rates and which are not  allowed to overlap. Now, the transcendental
relationship between PLM and the  granular compaction experiment is
easier to explain: both PLM and our coupled  flip-flop model belong to
the same {\em generic} class of AF-type systems (in  the case of PLM,
a local two-state mode is represented by a  particle whose  center of
mass may or may not be placed at point {\bf x}; the mode  coupling is
due to the hard-core interactions). The PLM is known to capture   the
slow dynamics of granular compaction and some features of its
fluctuation spectrum~\cite{comp2,alex}. In fact, we performed
numerical simulations of the PLM that display~\cite{josserand} the
memory effects  discussed in this work. 

In conclusion, we used a sequence of abrupt  switches  of the shaking
intensity $\Gamma$ to study the response of a vibrated granular
system. This technique can be used in the vicinity of the steady state
density, as  well as far from equilibrium. The major result is the
direct demonstration of a memory effect in the system:  the evolution
is not  predetermined by the local density alone, and its  description
requires introduction of additional ``hidden''  variables. Our
phenomenological model for this behavior is built on minimal
assumptions about the dynamics of the system and produces results
which are generic and are expected to be valid for a wide class of
more realistic microscopic models.

Acknowledgments: We would like to thank Sue Coppersmith, Leo Kadanoff,
Sid Nagel, and Tom Witten for insightful discussions and Fernando
Villarruel for help with the experiments.  This work was supported by
the  NSF   under Award CTS-9710991 and by the MRSEC Program of the NSF
under Award DMR-9808595.

\end{document}